# The emerging commercial landscape of quantum computing

Evan R. MacQuarrie[1,2,3,*], Christoph Simon[4,5], Stephanie Simmons[1,2], and Elicia Maine[3]
[1]Physics Department, Simon Fraser University, Canada
[2]Photonic Inc., Canada
[3]Beedie School of Business, Simon Fraser University, Canada
[4]Department of Physics and Astronomy, University of Calgary, Canada
[5]Institute for Quantum Science and Technology, University of Calgary, Canada
*e-mail: emacquar@sfu.ca

*Quantum computing technologies are advancing, and the class of addressable problems is expanding. Together with the emergence of new ventures and government-sponsored partnerships, these trends will help lower the barrier for new technology adoption and provide stability in an uncertain market. Until then, quantum computing presents an exciting testbed for different strategies in an emerging market.*

Despite technological advances and a wave of investment, the emerging quantum computing (QC) commercial market still faces a high level of both technological and market uncertainty. The opportunity present in this technological uncertainty has led to a rapid growth in the number of active QC ventures, and as research intensity increases to address outstanding technical challenges, a variety of business strategies have also emerged to tackle the market uncertainty. We examine the recent growth in the commercial QC market through the lens of dominant product design [1], and we contrast emerging strategies for developing the QC market.

**Market evolution**

The emergence of a new market is often rife with uncertainty, and predictors of how such markets will evolve have been heavily studied. The dominant design model of market evolution holds that the number of firms active in a sector provides a strong indicator of where the market sits in its life cycle [1-3]. The number of firms increases as ventures are drawn to the opportunity present in the technological uncertainty. Once a dominant design emerges, however, the number of firms shrinks through a process of consolidation and exit.

From the number of incumbent firms and private start-ups active in the field (Fig. 1a), we see that over the past two decades incumbent firms have been laying the foundations for commercializing QC with the number of start-up ventures lagging behind. However, the number of start-up companies began increasing in 2011 when D-Wave Systems sold its first quantum annealing system. By 2015 — one year before IBM's release of the first commercial cloud quantum computer — the number of start-ups surpassed the number of incumbents active in the field, and that number has been growing rapidly ever since.

The trend in Fig. 1a suggests that the incumbent firms' wealth of resources provided the extra stability necessary to endure the years of basic research that demonstrated commercial QC could be viable. As soon as that was demonstrated, however, private ventures began to flood the sector with new technologies and ambitions. This is an interesting demonstration of the idea that new ventures are more likely to pursue the commercialization of highly uncertain technologies. According to dominant design trends, we can expect that in the future standard practices will emerge, technical uncertainty will dissipate, and there will eventually be fewer market entries and a consolidation of firms [2,3]. For the time being, however, the opportunity in QC and the uncertainty behind which designs will become dominant continue to motivate new ventures to enter the industry.

An interesting variant of this trend appears in the number of firms actively involved in developing what we have termed QC software (Fig 1b), which include algorithms, applications, simulators, and interfaces. These technologies require far less capital to develop, which allowed software firms to proliferate once commercial cloud QC could provide a platform on which to develop their technologies.

**Patent trends**

Further insight into this rapid expansion of the quantum industry can be gained by looking at the number of patent applications by different firms across the globe. Following the search methodology detailed in Box 1, we plotted the number of patent applications by incumbent firms in the North American (Fig. 1c) and Asian (Fig. 1d) markets and those of several start-up firms (Fig. 1e). These plots reveal a stark contrast between Asian and American firms, where the majority of Asian incumbents have steadily grown their patent portfolios over two decades of developmental research. Apart from HP's early entrance and subsequent exit from the market, the American incumbents emerged much later but developed faster after IBM's 2016 commercial release. The Asian incumbents have not mirrored this reaction to the release of a commercial QC platform. With D-Wave Systems as the notable exception, start-up ventures across the globe have only recently emerged on the market but are rapidly patenting technologies to stake out their competitive advantage.

**Emergent strategies**

Despite this boom, the total number of users on commercial platforms remains [a small fraction](#) of the expected market. As efforts to advance the technology continue, firms have simultaneously expanded their push to validate their technologies and grow the QC market. As with other breakthrough technologies, governments have also played a significant role in funding the basic and applied research enabling the emergent industry. Governments around the world continue to make significant [investments](#) in the technological developments behind quantum hardware and software advances. Innovation policy can also support the development of emergent business strategies. For example, the National Science Foundation (NSF) recently [announced](#) a coordinated program with cloud QC providers IBM, Amazon Web Services, and Microsoft Azure. Under this program, the NSF will provide additional funding for graduate students and these providers will offer those students free access to their quantum cloud. Within this triumvirate of North American quantum companies, we have observed the emergence of two markedly different strategies for managing the QC value chain: the full-stack approach and structural open innovation.

**Full-stack approach**

IBM joins the NSF agreement with the richest pedigree in QC. From their early entry into the field with one of the first experimental executions of a quantum algorithm [4] in 1998 to their current fleet of seventeen quantum computers hooked up to the cloud, IBM has been a dominant presence in the community from the beginning. This has allowed them to pursue a `full-stack' business strategy wherein they produce the entire QC value chain. This value chain consists of the quantum hardware that executes the desired algorithms, classical hardware for controlling the quantum processor and connecting the processor to the cloud, software tools for interfacing a user with the cloud system, and a portfolio of applications and knowledge that a user can access.

IBM's strategy aims to fulfill each of these steps while also providing advanced classical simulators for algorithm development on simulated quantum computers. This approach has materialized into their IBM Q Network, a tiered community of companies and research institutions attempting to build the QC community through collaborative R&D and user open innovation [5]. This network aims to support, train, and educate users of IBM's cloud quantum computers while cooperatively developing practical applications for these new tools. Members of the network run the gamut from national laboratories and universities to Fortune 500 companies and QC software start-ups. The IBM Q Network does not contain any QC hardware start-ups or any direct competitors to IBM's qubit technology. In fact, only one member of the community — [advanced materials company Archer Materials Limited](#) — is actively developing competing QC hardware, and Archer's hardware pursuit is a moonshot room temperature hardware approach that is not directly competing with IBM's qubit technology.

**Structural open innovation**

In contrast to IBM's full-stack approach, Microsoft and Amazon have opted to pursue structural open innovation [5]. They have launched partnerships to integrate hardware and software quantum start-ups into a packaged value chain that they will deliver to users through their established cloud platforms. Amazon has partnered with hardware start-ups IonQ, Rigetti Computing, and D-Wave Systems, each representing a different qubit technology. Similarly, Microsoft has partnered with IonQ and Honeywell, which are both utilizing trapped ions qubits, and Quantum Circuits, Inc which is developing superconducting qubits. These partnerships supplement the global research network that Microsoft has built to pursue topological QC and their venture investment into the QC start-up PsiQuantum. Amazon has similarly signaled interest in launching their own internal hardware development program through a recent partnership with Caltech. By integrating both internal and external efforts, Microsoft and Amazon have been able to diversify their investment in different qubit architectures, which could prove beneficial if superconducting qubits hit a barrier that prevents further scaling-up. This strategy of structural open innovation has simultaneously allowed them to jump-start their QC offering and leverage their existing cloud computing base to begin a process of user open innovation.

Amazon and Microsoft's structural open innovation creates a symbiotic approach to market entrance that simultaneously grants start-up companies such as Rigetti and IonQ access to the customers and reputation of well-established firms. With the development of so-called Noisy Intermediate-Scale Quantum (NISQ) machines, QC is evolving from a challenge in search of a solution to a solution in search of users, and market entrance will become an increasingly important hurdle for start-ups to overcome. Incumbent partnerships such as those Microsoft and Amazon have created could offer a ready-made solution to this challenge.

**Outlook**

The recent wave of entries into the QC market is not without historical precedent. When the classical supercomputing market was searching for an alternative to the traditional von Neumann architecture, a similar upswell in firms developing massively parallel computers arose [2]. Once a dominant massively parallel computer design emerged, however, the number of firms declined sharply through a process of consolidation and exit. The current QC upswell has been driven by over one hundred start-up companies entering the sector. As was previously seen in the emerging nanobiotechnology industry [3], the diverse range of approaches undertaken by these start-ups precedes the market selection of a dominant design.

**Related links**

Where Will Quantum Computers Create Value—and When? https://www.bcg.com/en-ca/publications/2019/quantum-computers-create-value-when.aspx
Enabling Quantum Computing Platform Access for National Science Foundation Researchers with Amazon Web Services, IBM, and Microsoft Quantum
https://www.nsf.gov/pubs/2020/nsf20073/nsf20073.jsp
US investment in future technologies https://www.whitehouse.gov/articles/trump-administration-investing-1-billion-research-institutes-advance-industries-future/
Archer signs agreement with IBM https://archerx.com.au//src/uploads/2020/05/20200505_Quantum-computing-agreement-with-IBM-ASX-Release.pdf
PsiQuantum joins Microsoft's M12 https://medium.com/m12vc/welcoming-psiquantum-to-the-m12-portfolio-56926b455a8f
Amazon announces partnership with Caltech https://eas.caltech.edu/engenious/16/chair_message

Becoming a quantum developer with Microsoft Quantum
https://cloudblogs.microsoft.com/quantum/2020/05/19/azure-quantum-preview-new-developer-training-learning-tools/

**Competing interests**
S.S. and E.R.M. are employees of the quantum computing start-up company Photonic Inc.

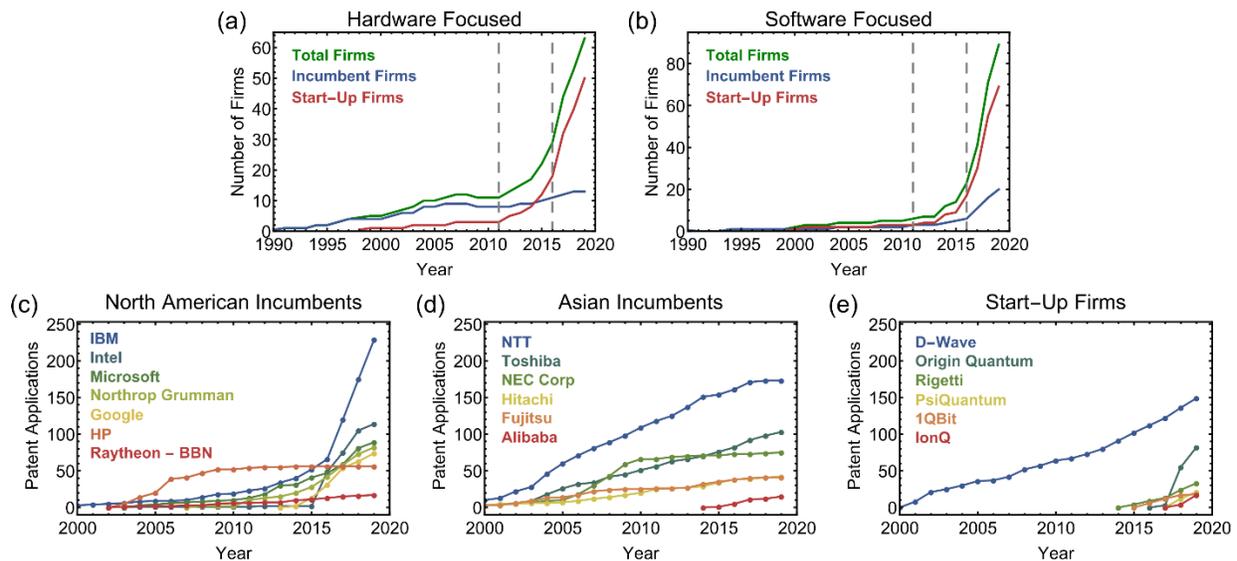

FIG. 1: Trends in commercial quantum computing. Number of incumbent and start-up companies developing quantum computing (a) hardware and (b) software over the past 30 years. The dashed lines at 2011 and 2016 indicate the years that D-Wave sold its first quantum annealing system and that IBM released the first commercial cloud quantum platform respectively. Patent applications worldwide for (c) North American incumbent firms, (d) Asian incumbent firms, and (e) start-up ventures over the past two decades. In (e), 1QBit is developing QC software; the other start-ups listed are hardware ventures.

**Box 1: Methodology**

For our dominant design analysis, a list of firms active in the sector was obtained using the database of companies listed on www.quantumcomputingreport.com, press releases, and patent databases. When a press release or a start (stop) date could not be found for a company, the first (last) patent application date was used as a start (end) date for that venture. We define a start-up company as a venture that has been established to develop quantum computing (QC) technology.

In capital-intensive markets with high uncertainty such as QC, patents are often used to stake out space or attract investors and potential partners. Our patent information was obtained using Google's patent database with a keyword search of "(`quantum computing' OR `quantum computer' OR `quantum information' OR `qubit?')". This search string developed by starting from ``quantum computing'', adding keywords, and monitoring which patents were added by each keyword. We tracked patent applications rather than patent grants to account for the youth of many firms in the sector. These applications were sorted by family to prevent double counting the same technology patented under multiple authorities, and a manual check of patent titles and descriptions was performed to eliminate any false positives. Firms with fewer than ten patent applications were then eliminated from the list, and the remaining firms were sorted into start-ups, North American, and Asian incumbents. Start-ups pursuing quantum communication technologies rather than QC that had made it through these filters were then eliminated from the list. This process provided us with a list of companies to include, but we found that our initial search had overlooked some patents. We then repeated this process for each remaining firm, including the company name in the search as the assignee. The results are plotted in Fig. 1c-e. Despite large QC efforts in both Europe and Australia, start-ups from these regions are not represented in Fig. 1e. This is because the European and Australian start-ups either have not been as active in patenting, are more focused on quantum communications than QC, or are partnered with corporations or universities who show up as the patent assignee.

It is worth noting that incumbent firms' efforts in quantum technologies often combine both quantum information and quantum communication pursuits. Our patent search methodology only offers a weak filter of these two, often-interleaved pursuits. Because of the large overlap between these efforts, this caveat does not alter the trends we report.